\documentclass[12pt,twoside]{article}

\begin{document}\hbadness=10000\thispagestyle{empty}
\pagestyle{myheadings}
\title{{\bf Quantum mechanics emerging from \\ ``timeless'' classical dynamics}} 
\author{$\ $\\
{\bf Hans-Thomas Elze} 
\\ $\ $\\ 
Instituto de F\'{\i}sica, Universidade Federal do Rio de Janeiro \\ 
C.P. 68.528, 21941-972 Rio de Janeiro, RJ, Brazil\footnote{E-mail address: thomas@if.ufrj.br}
}
\vskip 0.5cm
\maketitle
\vspace{-8.5cm}
\vspace*{8.0cm}
\begin{abstract}{\noindent
We study classical Hamiltonian systems in which the intrinsic 
proper time evolution parameter is related through a probability 
distribution to the physical time, which is assumed to be discrete.  
   In this way, a physical clock with discrete states is introduced, 
   which presently is still treated as decoupled from the system.   
This is motivated by the recent discussion of ``timeless'' 
reparametrization invariant models, where discrete physical time 
has been constructed based on quasi-local observables. Employing 
the path-integral formulation of classical mechanics developed 
by Gozzi et al., we show that these deterministic classical systems 
can be naturally described like unitary quantum mechanical models. We 
derive the emergent quantum Hamiltonian in terms of the underlying 
classical one. 
   Such Hamiltonians typically need a regularization -- 
   here performed by discretization -- in order to arrive at models 
   with a stable groundstate in the continuum limit.  
   This is demonstrated in two examples, recovering and generalizing 
   a model advanced by 't\,Hooft. 

\vskip 0.2cm
\noindent
PACS: 03.65.Ta, 04.20.-q, 05.20.-y 
}\end{abstract}
\section{Introduction}
Since its very beginnings, there has been a long series of speculations on the possibility of deriving 
quantum theory from more fundamental dynamical structures, possibly deterministic 
ones. Famous is the discussion by Einstein, Podolsky and Rosen. This lead to the 
EPR paradox, which in turn was interpreted by its authors as indicating the need for a more complete fundamental 
theory \cite{EPR}. However, just as numerous have been attempts to prove no-go theorems prohibiting 
exactly such ``fundamentalism''. This culminated in the studies of Bell, leading to the 
Bell inequalities \cite{Bell}. The paradox as well as the inequalities have come under experimental 
scrutiny in recent years, confirming the predictions of quantum mechanics in laboratory experiments on 
scales very large compared to the Planck scale.   
  
However, to this day, the feasible experiments cannot rule out the possibility that quantum mechanics emerges as an effective theory only on sufficiently large scales and can indeed  
be based on more fundamental models.     

Motivated by the unreconciled clash between general relativity and quantum theory, 't\,Hooft has 
strongly argued in favour of model building in this context \cite{tHooft01} (see also further 
references therein). He has shown in individual 
examples the emergence of the usual Hilbert space structure and unitary evolution in deterministic 
classical models in an appropriate large-scale limit. Particular emphasis has been placed on the 
observation that while it is relatively easy to arrive at a Hilbert space formulation for  
classical dynamics, it is difficult to obtain emergent Hamiltonians with a spectrum bounded 
from below, i.e., having a well-defined groundstate.   

Various further arguments for deterministically induced quantum features have recently been proposed, 
for example, in Refs.\,\cite{Wetterich02,Vitiello01,Smolin,Adler}, in the context of statistical systems, of  
considerations related to quantum gravity, and of matrix models, respectively.   
   
Our aim here is to contribute to this line of research by reporting a rather large class of 
classical deterministic systems which yield to a quantum mechanical description. This is based 
on our recent work on time-reparametrization invariant models \cite{ES02,E03}. In particular we 
have introduced the construction of a discrete physical time for such ``timeless'' systems, which will 
be the starting point of the present developments. In the following we briefly summarize essential aspects.   

{\it This paper does not claim to have derived quantum mechanics from an underlying classical 
dynamics yet. However, we develop a formalism which may prove useful for this purpose, besides  
having potentially other interesting applications in statistical mechanics. We do illustrate our 
approach with first examples indicating the possible emergence of quantum mechanics from deterministic 
models.}\footnote{A summary of this work was presented as invited talk at the International Workshop 
{\it Trends and Perspectives on Extensive and Non-Extensive Statistical Mechanics} 
in honor of Constantino Tsallis' 60th birthday, Angra dos Reis (Brazil), Nov.19-21, 2003 \cite{Angra}.}

\subsection{Discrete physical time in ``timeless'' classical models}  
Similarly to common gauge theories, such as those of the standard model of particle physics, 
reparametrization invariant systems show invariance under a kind of gauge transformations. 
In the most general case of diffeomorphism invariant theories, such as general relativity 
or string theory, this amounts to invariance under general coordinate transformations. 
  
We limit ourselves to time-reparametrization invariance here, which for our purposes can be 
expressed as invariance of the dynamics under arbitrary transformations:\footnote{Some restrictions 
are imposed in Refs.\,\cite{ES02,E03} on physical grounds, such as differentiability and monotonicity 
of the function $f$.} 
\begin{equation}\label{timerepara} 
t\;\longrightarrow t'\;,\;\;\mbox{with}\;\;t\equiv f(t') 
\;\;. \end{equation}    
Details of the corresponding constrained Lagrangian dynamics can be found in Refs.\,\cite{ES02,E03} 
for the respective models. 
 
Similarly as Gauss' law in electrodynamics, for example, a most important consequence of time-reparametrization 
invariance is a (weak) constraint, which states that the Hamiltonian has to vanish (on the solutions of the equations 
of motion).\footnote{In the case of the free relativistic particle the 
invariance amounts to invariance under reparametrization of the proper time parameter and the ensuing 
constraint is the familiar mass-shell constraint on its momentum.}    
Since the Hamiltonian commonly is the generator of time evolution of the system, this is 
what has led to name this type of systems ``timeless''. A Newtonian external time parameter does not exist. 
This becomes problematic when trying to quantize such systems, since a standard Schr\"odinger 
equation does not exist either. In particular, the  
Wheeler-DeWitt equation, $\widehat H|\psi\rangle =0$, epitomizes the intrinsic problems of quantum gravity, 
seen from this perspective.  
  
Numerous approaches have been tried to resolve this (in)famous ``problem of time''. For our purposes, 
it may suffice to mention one, in order to contrast it with our proposal. 

In Refs.\,\cite{Rovelli90,MRT99,Montesinos00} it has been assumed that {\it global}  
features of a suitably parametrized trajectory of the system, are accessible to the observer. This makes 
it possible, in principle, to express the evolution of an arbitrarily selected degree of freedom relationally 
in terms of others. Naively speaking, the question ``What time is it?'' is replaced by ``What is the value $x$ 
of observable $X$, {\it when} observable $Y$ has 
value $y$?''. Thereby the Hamiltonian and possibly additional constraints have been eliminated 
in favour of Rovelli's ``evolving constants of motion''. 
  
In distinction, we insist on a {\it local} description. 
We have shown that for a particle with 
time-reparametrization invariant dynamics, be it relativistic or nonrelativistic, 
one can define quasi-local observables 
which characterize the evolution in a gauge invariant way \cite{ES02,E03}.  
    
Essentially, we employ some of the degrees of freedom of the system to trigger a localized 
``detector''. This can be defined in an invariant way. It amounts to attributing to 
an observer the capability to count discrete events. -- In passing we remark that a similar 
approach, avoiding the notion of time altogether and replacing it by counts of idealized 
coincidence detectors in phase space has recently also been put forth by Halliwell and 
Thorwart \cite{Jonathan}. -- In any case, the detector counts present an observable measure 
of time. This {\it physical time is discrete}. 

This result can be understood in a different way, by noting that our construction   
is practically based on a Poincar\'e section, or subsection thereof, which reflects an ergodic 
if not periodic aspect of the dynamics -- quite analogous to a pendulum 
which triggers a coincidence counter each ``time'' it passes through its equilibrium 
position.   
Reparametrization invariance strongly limits the information which can be extracted from it  
with respect to a complete trajectory. This is the underlying reason that a physical time 
based on local observations (clock readings) necessarily is discrete.\footnote{This does not conflict with certain cosmological models where a scalar matter field coupled to 
gravity is invoked as a continuous time variable. Here, the need to 
know the scalar field globally can be circumvented by assuming 
its homogeneity.}  
  
Independently of our physical motivation of discrete time, we remark that 
the possibility of a fundamentally discrete time (and possibly other discrete 
coordinates) has been explored before, ranging from an early realization of Lorentz 
symmetry in such a case \cite{Snyder} to detailed explorations of its consequences 
and consistency in classical mechanics, quantum field theory, and general relativity 
\cite{TDL83,JN97,Pullin}. Further recent developments are discussed in the review of Ref.\,\cite{AmelinoC};      
particularly in relation to fundamental or induced violations of Lorentz symmetry, which 
are believed to come within experimental reach in the near future.    

So far, however, no classical physical models giving rise to such discreteness were proposed.  
{\it Quantization as an additional step} -- which results in discreteness of coordinates in 
various cases -- has always been performed as usual. 

\subsection{Does discrete time induce quantum mechanical features?}  
It is the purpose of the present study to reach a qualified `Yes', answering this question.  
We hope this will contribute to the study of potentially deterministic substructures 
leading to quantum mechanics as an emergent theory.
  
Based on the findings of Refs.\,\cite{ES02,E03}, respectively, we have argued there that 
those discrete-time models can be mapped on a cellular automaton studied 
by 't\,Hooft before \cite{tHooft01}. With the help of the algebra of $SU(2)$ generators, 
it has been shown that these models actually reproduce the quantum mechanical harmonic 
oscillator in a suitably defined large-scale limit. 
  
We will come back to variants of the earlier results in Section\,6.    
However, in the following main parts of this paper, we attempt to show more generally 
that due to inaccessability of globally complete information on trajectories of the system, 
the evolution of remaining degrees of freedom appears as in a quantum mechanical model 
when described in relation to the discrete physical time.     
  
We may call this ``stroboscopic'' quantization: when a continous 
physical time is not available but a discrete one is -- like reading an analog clock under 
a stroboscopic light -- then states of the system which fall in between subsequent clock 
``ticks'' cannot be resolved. (Of course, evolution in the unphysical parameter time 
is continous in the constrained Lagrangian models we refer to.) Such unresolved states  
form equivalence classes which can be identified with primordial 
Hilbert space states \cite{tHooft01,Wetterich02,ES02,E03}. The residual dynamics then has 
to describe the evolution 
of these states through discrete steps. Under favourable circumstances, this results in unitary 
quantum mechanical evolution, as we shall see.   

Our present aim is to show that this occurs quite generally in classical Hamiltonian systems, 
if time is discrete. We will presently simplify the situation by assuming 
that the physical time can be related by a probability distribution to the proper time of 
the equations of motion. Explicit examples of such behaviour can be found in Refs.\,\cite{ES02,E03}, 
when the clock degrees of freedom evolve independently of the rest of the system, apart 
from the Hamiltonian constraint.    

Thus, while the investigation of the coupled system-clock dynamics is 
presently under study \cite{E03a}, here we make the approximation that corresponding 
``backreaction'' effects are small. This feature, besides characterizing a good clock, 
may serve as an appropriate simplification in our exploratory steps.     
In the concluding section, we will briefly comment about  
extensions, where the prescribed probabilistic mapping of physical onto proper time 
shall be abandoned in favour of a selfconsistent treatment. A closed system, of course, has   
to include its own clock, if it is not entirely static, reflecting the experience  
of an observer in the universe. 
  
Finally, in order to put our approach into perspective, we remark that there is clearly no need to 
follow such construction leading to a discrete physical time in ordinary mechanical systems or 
field theories, where time is an external classical parameter. 
However, assuming for the time being that truly fundamental theories 
will turn out to be diffeomorphism invariant, adding further the requirement of 
the observables to be quasi-local,\footnote{For example, a closed string loop 
representing a fundamental length might define the ultimate resolution of distance 
measurements and, thus, condition the notion of locality.} when describing the evolution, then 
such an approach seems natural, which promises to lead to quantum mechanics as an 
emergent description or effective theory on the way.     

The paper is organized as follows. Section\,2 provides a selfcontained brief summary of 
the path-integral formulation of classical mechanics. We will employ this as a convenient tool 
to formulate our approach. In later developments one might also introduce a cooresponding 
operator formalism, for example, considering as a starting point 
the early work of Koopman and von\,Neumann \cite{KN}. 
  
In Section\,3 the relevant to-be-quantum states are introduced, i.e. equivalence classes of 
states of the underlying classical system, as we mentioned. Their evolution 
is studied in Section\,4, leading to an emergent Hamilton operator and associated Schr\"odinger 
equation under circumstances to be discussed there. In Section\,5 the calculation of physical 
time dependent observables, 
or expectation values of corresponding operators, is related to the states. 

Section\,6 presents 
some simple examples of the emergent Hamiltonians and the calculation of their spectra. 
We show that -- under the present simplifying assumptions -- these operators 
need to be regularized, in order to represent acceptable quantum models, with a stable 
groundstate in particular. Here we achieve this by discretization. We speculate that 
its apparent arbitrariness might be removed by a future selfconsistent 
treatment of clock-system interactions, which will lead to dissipative effects on 
small scales and could define a unique quantum system for each classical model with 
discrete time. The concluding Section\,7 presents a brief summary of the presented 
work and points out some open problems.    

\section{Classical mechanics via path-integrals}   
Classical mechanics can be cast into path-integral form, as originally 
developed by Gozzi, Reuter and Thacker \cite{GRT}, and with recent addenda reported in Ref.\,\cite{GR00}.  
While the original motivation has been to provide a better understanding of geometrical aspects of quantization,   
we presently use it as a convenient tool. We refer the interested reader to the cited references for details,  
on the originally resulting extended (BRST type) symmetry in particular. Here we suitably incorporate 
time-reparametrization invariance, assuming equations of motion written 
in terms of proper time (as in our earlier examples \cite{ES02,E03}).
  
Let us begin with a $(2n)$-dimensional classical phase space ${\cal M}$ with coordinates denoted 
collectively by $\varphi^a\equiv (q^1,\dots ,q^n;p^1,\dots ,p^n),\;a=1,\dots ,2n$, where 
$q,p$ stand for the usual coordinates and conjugate momenta. Given the proper-time independent 
Hamiltonian $H(\varphi )$, the equations of motion are: 
\begin{equation}\label{eom}
\frac{\partial}{\partial\tau}\varphi^a=\omega^{ab}\frac{\partial}{\partial\varphi^b}H(\varphi ) 
\;\;, \end{equation} 
where $\omega^{ab}$ is the standard symplectic matrix and $\tau$ denotes the proper time; 
summation over indices appearing twice is understood.      
 
To the equation of motion we add the (weak) Hamiltonian constraint, $C_H\equiv H(\varphi )-\epsilon 
\simeq0$, with $\epsilon$ a suitably chosen parameter. This constraint has to be satisfied by the solutions 
of the equations of motion. Generally,  
it arises in reparametrization invariant models, similarly as the mass-shell constraint in the case  
of the relativistic particle \cite{E03}. It is necessary when the  
Lagrangian time parameter is replaced by the proper time in the equations 
of motion. In this way, an arbitrary ``lapse function'' is eliminated, which 
otherwise acts as a Lagrange multiplier for this constraint. 
 
We remark that field theories can be treated analogously, considering indices $a,b$, etc. 
as continuous variables.  
  
Starting point for our following considerations is the {\it classical} generating 
functional, 
\begin{equation}\label{Z} 
Z[J] \equiv \int_H{\cal D}\varphi\;\delta [\varphi^a(\tau )-\varphi^a_{cl} (\tau )]\exp (i\int\mbox{d}\tau\;J_a\varphi^a) 
\;\;, \end{equation} 
where $J\equiv\{ J_{a=1,\dots ,2n}\}$ is an arbitrary external source, 
$\delta [\cdot ]$ denotes a Dirac $\delta$-functional,  
and $\varphi_{cl}$ stands for a solution of the classical 
equations of motion satisfying the Hamiltonian constraint; its presence is indicated by the subscript ``$H$'' 
on the functional integral.  
It is important to realize that $Z[0]$, as it stands, gives weight 1 to any classical path satisfying the constraint 
and zero otherwise, integrating over all initial conditions.   
  
Using the functional equivalent of $\delta (f(x))=|\mbox{d}f/\mbox{d}x|_{x_0}^{-1}\cdot\delta (x-x_0)$\,,  
the $\delta$-functional under the integral for $Z$ can be replaced according to: 
\begin{equation}\label{delta} 
\delta [\varphi^a(\tau )-\varphi^a_{cl} (\tau )]=>
\delta [\partial_\tau\varphi^a-\omega^{ab}\partial_bH]
\det [\delta^a_b\partial_\tau -\omega^{ac}\partial_c\partial_bH] 
\;\;, \end{equation} 
slightly simplifying the notation, e.g. $\partial_b\equiv\partial /\partial\varphi^b$. 
Here the modulus of the functional determinant has been dropped \cite{GRT,GR00}.   

Finally, the $\delta$-functionals and determinant are  
exponentiated, using the functional Fourier representation and ghost variables, respectively. 
Thus, we obtain the generating functional in the convenient form: 
\begin{equation}\label{Zexp} 
Z[J]=\int_H{\cal D}\varphi{\cal D}\lambda{\cal D}c{\cal D}\bar c 
\;\exp \Big (i\int\mbox{d}\tau (L+J_a\varphi^a)\Big )
\;\;, \end{equation} 
which we abbreviate as 
$Z[J]=\int_H{\cal D}\Phi\;\exp (i\int\mbox{d}\tau L_J)$. 
The enlarged phase space is $(8n)$-dimensional, consisting of points described by the 
coordinates $(\varphi^a,\lambda_a,c^a,\bar c_a)$. The 
effective Lagrangian is now given by \cite{GRT,GR00}: 
\begin{equation}\label{L} 
L\equiv\lambda_a\Big (\partial_\tau\varphi^a-\omega^{ab}\partial_bH\Big ) 
+i\bar c_a\Big (\delta^a_b\partial_\tau -\omega^{ac}\partial_c\partial_bH\Big )c^b
\;\;, \end{equation} 
where $c^a,\bar c^a$ are anticommuting Grassmann variables. 
We remark that an entirely bosonic version of the path-integral exists \cite{GR00}.  

This completes our brief review of how to put (reparametrization invariant) 
classical mechanics into path-integral form.

\section{From discrete time to ``states''}   
We recall from the motivation provided in Section\,1.1 that discrete physical time $t$ 
is constructed and based on the counting of suitably defined incidents. In particular, 
we have coincidences in mind, ``when'' points belonging to the trajectory of the system 
coincide with the position of an idealized detector. For concrete realizations of this 
procedure and study of its invariance as well as further intrinsic properties we refer the 
interested reader to Refs.\,\cite{ES02,E03}, and to Ref.\,\cite{Jonathan} for a similar 
construction. 

Thus, physical time is measured by a nonnegative integer multiple of some 
unit time, $t\equiv nT$, and is read off from a sort of localized track counting device.\footnote{In 
our earlier detailed examples we were motivated by the at present actively researched separation in 
higher dimensional (cosmological) models of bulk and brane degrees of freedom: interactions 
between both kinds of matter could lead to a more detailed picture of such countable incidents 
providing the basis of physical time.} For our present study details of the clock construction 
are of secondary importance. Rather we investigate the consequences of the discreteness of 
physical time. In particular, we want to demonstrate the existence of quantum mechanical 
features which may derive from it.  
       
In any case, then, the proper time $\tau$ parametrizing the evolution should be calculable 
in terms of $t$. As can be observed, for example, in our earlier numerical simulations, this 
can be a formidable task, depending on the dynamics governing the underlying classical 
system \cite{ES02,E03}. Therefore, an analytic approach for interacting clock-coupled-to-system 
models is beyond the scope of this work and considered elsewhere \cite{E03a}.      

We will assume that the backreaction and memory effects which generally result from 
the coupling between clock and system are small and negligible for our present purposes.  
Then the relation between the discrete physical time $t$ and the proper time $\tau$ 
of the equations of motion, Eq.\,(\ref{eom}), can be represented by a 
time independent normalized probability 
distribution $P$: 
\begin{equation}\label{P} 
P(\tau ;t)\equiv P(\tau -t)\equiv\exp\Big (-S(\tau -t)\Big )\;\;,\;\;\;
\int\mbox{d}\tau\;P(\tau ;t)=1
\;\;. \end{equation}  
Note, in particular, that the perfect clock described in this way does not age with 
physical time. Eventually, we will also invoke the limiting case,  
$P(\tau ;t)=\delta (\tau -t)$, i.e. a deterministic mapping between $t$ and $\tau$. 

We remark that in the present situation, the Hamiltonian constraint needs to be applied 
to the system degrees of freedom only, while generally clock plus system are constrained as a whole.    
This is exemplified in all detail in our toy models \cite{ES02,E03}. 

Correspondingly, we introduce the modified   
generating functional:   
\begin{equation}\label{Zdef}
Z[J]\equiv\int_H\mbox{d}\tau_i\mbox{d}\tau_f
\int{\cal D}\Phi\;\exp\Big (
i\int_{\tau_i}^{\tau_f}\mbox{d}\tau\; L_J-S(\tau_i-t_i)-S(\tau_f-t_f)\Big ) 
\;\;, \end{equation} 
instead of Eq.\,(\ref{Zexp}), and using the condensed notation introduced there. In the present case, 
$Z[0]$ sums over all classical paths satisfying the 
constraint with weight $P(\tau_i;t_i)\cdot P(\tau_f;t_f)$, depending on their initial 
and final proper times, while all other paths get weight zero. 
In this way, 
the distributions of proper time values $\tau_{i,f}$ associated with the initial and 
final physical times, $t_i$ and $t_f$, respectively, are incorporated. 

Next, we insert $1=\int\mbox{d}\tau P(\tau ;t)$ into the expression for $Z$, 
with an arbitrarily chosen physical time $t_i<t<t_f$.   
This leads us to factorize 
the path-integral into two connected ones: 
\begin{eqnarray} 
Z[J]=\int\mbox{d}\tau\;P(\tau ;t)
&\cdot&\int\mbox{d}\tau_f\int_H{\cal D}\Phi_>\;\exp\Big (
i\int_{\tau}^{\tau_f}\mbox{d}\tau'\; L_J^>-S(\tau_f-t_f)\Big ) 
\nonumber \\ [1ex] 
&\cdot&\int\mbox{d}\tau_i\int_H{\cal D}\Phi_<\;\exp\Big (
i\int_{\tau_i}^{\tau }\mbox{d}\tau''\; L_J^<-S(\tau_i-t_i)\Big ) 
\nonumber \\ [1ex]\label{Zfactors}
&\cdot&\prod_a\delta(\varphi^a_>(\tau )-\varphi^a_<(\tau ))
\;\;,  \end{eqnarray} 
where ``$<$'' and ``$>$'' refer to earlier and later than $\tau$, respectively. 

The ordinary $\delta$-functions assure 
continuity of the classical paths in terms of the coordinates $q^a,\;a=1,\dots ,n$   
and momenta $p^a,\;a=1,\dots ,n$, at the (distributed) proper time $\tau$. This is necessary, 
in order to avoid a double-counting as compared to the original expression for $Z$, 
Eq.\,(\ref{Zdef}). Since any given classical path beginning at $t_i$ and ending at $t_f$ (with associated values 
$\tau_{i,f}$) is cut in two parts, without the continuity condition, the continuing 
part (``$>$'') would again have arbitrary initial conditions at the instant $\tau$, unlike the 
corresponding original path where this is excluded by the Hamiltonian flow. 
To put it differently, without continuity, originally $N$ independent paths would be factorized 
erroneously into $N^2$ independ ones.  

Another remark is in order here. Due to the presence of the Hamiltonian 
constraints on both parts of the cut trajectory, one of the $\delta$-functions is redundant. The 
resulting $\delta (0)$ may eventually be absorbed into the normalization of the states, 
which will be introduced now. 

Exponentiating the $\delta$-functions via Fourier transformation, the 
generating functional can indeed be interpreted as a scalar product of a state and 
its adjoint:  
\begin{equation}\label{Zstates}
Z[J]=\int\mbox{d}\tau\mbox{d}\pi\;P(\tau )\langle t;t_f|\tau ,\pi\rangle\langle \tau ,\pi |t;t_i\rangle 
\;\;, \end{equation} 
in ``$\tau ,\pi$-representation''; here $\mbox{d}\pi\equiv\prod_a(\mbox{d}\pi_a/2\pi )$. 
The {\it state} is defined by the path-integral: 
\begin{equation}\label{state} 
\langle\tau ,\pi_a|t;t_0\rangle\equiv 
\int\mbox{d}\tau'\int_H{\cal D}\Phi\;\exp\Big (
i\int_{\tau'}^{\tau +t}\mbox{d}\tau''L_J
-S(\tau'-t_0)+i\pi_a\varphi^a(\tau +t)\Big ) 
\;\;, \end{equation} 
and, similarly, the {\it adjoint state}: 
\begin{equation}\label{adjstate} 
\langle t;t_0|\tau ,\pi_a\rangle\equiv  
\int\mbox{d}\tau'\int_H{\cal D}\Phi\;\exp\Big (
i\int_{\tau +t}^{\tau'}\mbox{d}\tau''L_J
-S(\tau'-t_0)-i\pi_a\varphi^a(\tau +t)\Big ) 
\;\;, \end{equation} 
where $t_0$ represents an arbitrary physical time with which the respective paths 
begin, in the case of states, and end, in the case of adjoint states.  

We observe a redundancy in designating the states, which depend on the sum 
of proper and physical time only. This arises by the shift of $\tau$ leading to Eq.\,(\ref{Zstates}), 
noting that the probability distribution $P$ is not explicitly depending on the physical time, as we discussed. 

Furthermore, there is a symmetry between states and adjoint states: 
\begin{equation}\label{ccstates} 
\langle t;t_0|\tau ,\pi\rangle =\langle\tau ,\pi |t;t_0\rangle^\ast   
\;\;, \end{equation} 
if referred to the same reference time $t_0$, which is familiar in Hilbert space.  

Finally, the scalar product of any two states can be defined by: 
\begin{equation}\label{overlap}
\langle t_2;t_f|t_1;t_i\rangle\equiv Z[J]^{-1} 
\int\mbox{d}\tau\mbox{d}\pi\;P(\tau )\langle t_2;t_f|\tau ,\pi\rangle\langle\tau ,\pi |t_1;t_i\rangle 
\;\;. \end{equation}
Of course, we have $\langle t;t_f|t;t_i\rangle =1$, corresponding to Eq.\,(\ref{Zfactors}),  
while for $t_2\neq t_1$, the states generally are not orthogonal.      

Closing this section, we point out that here the path-integrals have always been 
considered to include all classical paths. However, they can be restricted by 
imposing (final) initial conditions in the case of (adjoint) states. Since this has no 
effect on the following study of the time evolution of a  
generic state, there is presently no need to make explicitly use of this.    

\section{Unitary evolution}   
In order to learn about the time evolution of generic states 
(in the absence of a source, $J=0$), we proceed similarly as in 
Eqs.\,(\ref{Zdef})--(\ref{state}).     

Suitably inserting ``1'', as before, and splitting the path-integral (at time $t$) which 
defines a state (at time $t'$), we obtain:  
\begin{equation}\label{evol} 
\langle\tau',\pi'|t'\rangle=\int\mbox{d}\tau\mbox{d}\pi\;P(\tau) 
\langle\tau',\pi'|\widehat U(t',t)|\tau ,\pi\rangle\langle\tau ,\pi |t\rangle
\;\;, \end{equation}
where henceforth we suppress to indicate the reference time $t_0$, see Eq.\,(\ref{state}), 
in order to simplify the notation. 
Written as a matrix element of an evolution operator $\widehat U(t',t)$, the kernel  
which appears is:    
\begin{equation}\label{kernel} 
\langle\tau',\pi'|\widehat U(t',t)|\tau ,\pi\rangle
\equiv\int_H{\cal D}\Phi\;\exp\Big (
i\int_{\tau +t}^{\tau'+t'}\mbox{d}\tau''L
+i\pi'\cdot\varphi (\tau'+t')
-i\pi\cdot\varphi (\tau +t) 
\Big ) 
\;\;, \end{equation}
with the functional integral over {\it all paths} running between $\tau +t$ and $\tau'+t'$, 
subject to the Hamiltonian constraint; 
we abbreviate: $\pi\cdot\varphi\equiv\pi_a\varphi^a$.  
  
Then, first of all, it is straightforward to establish the following composition rule: 
\begin{equation}\label{composition}
\widehat U(t'',t')\cdot\widehat U(t',t)=\widehat U(t'',t) 
\;\;, \end{equation} 
written here in the way of a matrix product which is to be 
interpreted as integration over intermediate variables, say $\mbox{d}\tau',\mbox{d}\pi'$\,, 
with appropriate weight factor $P(\tau')$. Then, 
the first integration, say over $\mbox{d}\tau'$, effectively removes a ``1'' -- 
such as we inserted before, in order to factorize path-integrals.  
The second, say over $\mbox{d}\pi'$ reinstitutes $\delta$-functions -- 
such as those which first made their appearance in Eq.\,(\ref{Zfactors}) -- which 
serve to link the endpoint coordinates of one classical path to the 
initial of another. This leads to the result of the right-hand side.  

Since the Hamiltonian constraint is a constant of motion, there is no need to 
constrain the path-integral representing the evolution operator. Integrating over 
intermediate variables removes all contributions violating the Hamiltonian 
constraint, provided we work with properly constrained states. This will be 
further discussed in the following section. 
    
We observe in Eq.\,(\ref{kernel}) that the physical-time dependence of the evolution operator (on $t'$ and $t$)  
amounts to translations of proper time variables ($\tau'+t'$ and $\tau +t$) in the matrix elements. 
This simplicity, of course, is related  
to the analogous property of the states, see Eq.\,(\ref{state}), which we discussed in 
the previous Section\,3. 

Tracing the steps backwards which led from Eq.\,(\ref{Z}) to Eq.\,(\ref{Zexp}) in Section\,2, 
we can similarly rewrite here the functional integral of Eq.\,(\ref{kernel}):   
\begin{equation}\label{Ut1}
\langle\tau',\pi'|\widehat U(t',t)|\tau ,\pi\rangle
=\int{\cal D}\varphi\;\delta [\varphi^a(\tilde\tau )-\varphi^a_{cl}(\tilde\tau )]
\exp\Big (i\pi'\cdot\varphi (\tau'+t')-i\pi\cdot\varphi (\tau +t)\Big )
\;\;, \end{equation} 
where the paths parametrized by $\tilde\tau$ run between $\tau +t$ and $\tau'+t'$ and all 
initial conditions are integrated over. 

Fixing momentarily the initial condition of the paths, $\varphi (\tau +t)=\varphi_i$, 
we select a particular path contributing in Eq.\,(\ref{Ut1}). By integrating over all 
initial conditions in the end, $\int\mbox{d}\varphi_i$, we recover the full expression. 
Again, this amounts to a suitable insertion of ``1'', a familiar procedure by now,  
which allows us to manipulate the above equation as follows:  
\begin{eqnarray}\label{Ut2} 
&\;&\langle\tau',\pi'|\widehat U(t',t)|\tau ,\pi\rangle 
\\ [1ex] 
&=&\int{\cal D}\varphi\int\mbox{d}\varphi_i\;\delta (\varphi (\tau +t)-\varphi_i)\; 
\delta [\varphi^a(\tilde\tau )-\varphi^a_{cl}(\tilde\tau )]
\exp\Big (i\pi'\cdot\varphi (\tau'+t')-i\pi\cdot\varphi (\tau +t)\Big )
\nonumber \\ [1ex] 
&=&\int\mbox{d}\varphi_i\;
\exp\Big (i\pi'\cdot\varphi_f-i\pi\cdot\varphi_i\Big )
\int{\cal D}\varphi\;\delta [\varphi^a(\tilde\tau )-\varphi^a_{cl}(\tilde\tau )]
\nonumber
\;\;, \end{eqnarray} 
where $\varphi_f$ denotes the value of $\varphi (\tilde\tau )$ at the endpoint $\tau'+t'$ 
of the path singled out by the initial condition $\varphi (\tau +t)=\varphi_i$. 
This value is determined by the classical equations of motion and will be calculated 
shortly. Since the remaining functional integral equals one -- only the one path with 
fixed initial conditon and weight 1 is contributing -- we first obtain the intermediate 
result:  
\begin{equation}\label{intermed} 
\langle\tau',\pi'|\widehat U(t',t)|\tau ,\pi\rangle 
=\int\mbox{d}\varphi_i\;
\exp\Big (i\pi'\cdot\varphi_f-i\pi\cdot\varphi_i\Big )
\;\;, \end{equation} 
where $\mbox{d}\varphi_i\equiv\prod_{a=1}^{2n}\mbox{d}\varphi^a_i$. 
Of course, the time dependent relation between $\varphi_f$ and $\varphi_i$ has 
to be enforced, in order to make this condensed expression explicit. This we do next. 

We employ the Liouville operator, 
\begin{equation}\label{Liouville}
\widehat{\cal L}\equiv -\frac{\partial H}{\partial\varphi}\cdot\omega\cdot
\frac{\partial}{\partial\varphi}
\;\;, \end{equation} 
where $\omega$ is the symplectic matrix, stemming from the Hamiltonian equations 
of motion (\ref{eom}). As is well known, the Liouville operator allows to generate  
a classical solution of these equations at any finite 
time, starting with a given initial condition. Applying this for our purposes yields:     
\begin{equation}\label{classprop}  
\varphi_f\equiv\varphi (\tau'+t')=\exp [\widehat{\cal L}(\tau'+t'-\tau -t)]
\varphi (\tau +t)\equiv\exp [\widehat{\cal L}(\tau'+t'-\tau -t)]
\varphi_i 
\;\;. \end{equation} 
Then, inserting into Eq.\,(\ref{intermed}), we obtain the simple but central result:  
\begin{eqnarray}\label{Ut3}
\langle\tau',\pi'|\widehat U(t',t)|\tau ,\pi\rangle 
&=&\int\mbox{d}\varphi\;
\exp\Big (i\pi'\cdot (\exp [\widehat{\cal L}(\tau'+t'-\tau -t)]
\varphi )-i\pi\cdot\varphi \Big )
\\ [1ex] \label{Ut4} 
&\equiv&{\cal E}(\pi',\pi ;\tau'+t'-\tau -t) 
\;\;, \end{eqnarray} 
where the by now superfluous subscript ``$i$'' has been omitted. 

Using Eq.\,(\ref{Ut3}), one readily confirms Eq.\,(\ref{composition}) once again. 
In particular, then $\widehat U(t|t')\cdot\widehat U(t'|t)=\widehat U(t|t)$, 
which is not diagonal, in general, in this $\tau,\pi$-representation. We have:  
$U(\tau',\pi';t|\tau ,\pi ;t)={\cal E}(\pi',\pi ;\tau'-\tau )$.  
  
In order to proceed, we consider the time dependence of the evolution kernel ${\cal E}$.  
Beginning with Eq.\,(\ref{Ut4}), one derives the equation: 
\begin{eqnarray}
i\partial_\tau {\cal E}(\pi',\pi ;\tau )&=&
-\int\mbox{d}\varphi\; 
\exp \Big (i\pi'\cdot\varphi (\tau )-\pi\cdot\varphi\Big )\;\pi'\cdot (\partial_\tau\varphi (\tau ))  
\nonumber \\ [1ex] 
&=&-\int\mbox{d}\varphi\; 
\exp \Big (i\pi'\cdot\varphi (\tau )-\pi\cdot\varphi\Big )\;\pi'\cdot\omega\cdot
\frac{\partial}{\partial\varphi}H(\varphi (\tau )) 
\nonumber \\ [1ex] \label{Et} 
&=&\widehat{\cal H}(\pi',-i\partial_{\pi'}){\cal E}(\pi',\pi ;\tau )
\;\;, \end{eqnarray} 
with the effective {\it Hamilton operator}: 
\begin{equation}\label{Hamiltonian} 
\widehat{\cal H}(\pi ,-i\partial_\pi )\equiv -\pi \cdot\omega\cdot
\frac{\partial}{\partial\varphi}H(\varphi )|_{\varphi =-i\partial_{\pi}}  
\;\;. \end{equation} 
Here we also used Eq.\,(\ref{classprop}) in the first step, the equations of motion 
(\ref{eom}) in the second, and suitably pulled the factor following the exponential out 
of the integral at last. The initial condition,   
\begin{equation}\label{Einitial} 
{\cal E}(\pi',\pi ;0)=(2\pi )^{2n}\delta^{2n}(\pi'-\pi) 
\;\;, \end{equation} 
can be read off from Eq.\,(\ref{Ut3}).   
Integrating Eq.\,(\ref{Et}), we immediately obtain: 
\begin{equation}\label{Esol}
{\cal E}(\pi',\pi ;\tau )
=(2\pi )^{2n}\exp [-i\tau \widehat{\cal H}(\pi' ,-i\partial_{\pi'})]
\delta^{2n}(\pi'-\pi) 
\;\;, \end{equation} 
taking the initial condition into account.  

This result, which yields the evolution operator $\widehat U$ by Eq.\,(\ref{Ut4}), 
finally allows to describe the evolution of a generic time dependent state,  
$|\Psi (t)\rangle$, which takes place in one discrete physical time step 
(unit $T$). Using Eq.\,(\ref{evol}), we calculate:  
\begin{eqnarray}
\langle\tau',\pi'|\Psi (t+T)\rangle &=&\int\mbox{d}\tau\mbox{d}\pi\;P(\tau) 
{\cal E}(\pi',\pi ;\tau'+T-\tau )
\langle\tau ,\pi |\Psi (t)\rangle
\nonumber \\ [1ex] \label{discrSchroed}
&=&\int\mbox{d}\tau\;P(\tau) 
\exp [-i(\tau'+T-\tau )\widehat{\cal H}(\pi' ,-i\partial_{\pi'})]
\langle\tau ,\pi'|\Psi (t)\rangle
\;\;, \end{eqnarray}
where the integration over $\mbox{d}\pi\equiv\prod_a(\mbox{d}\pi_a/2\pi )$ has been carried 
out with the help of the 
$\delta$-functions of Eq.\,(\ref{Esol}). This equation 
appears like a {\it discrete time Schr\"odinger equation} and 
presents our main result of this section. However, some qualifying 
remarks are due here.     

\subsection{Discussion}   
In order to facilitate the investigation of the properties of Eq.\,(\ref{discrSchroed}), 
let us assume that a generic time dependent state shares the following 
(proper) time translation property  
with the state defined in Eq.\,(\ref{state}): 
\begin{equation}\label{ttransl} 
\langle\tau ,\pi |\Psi (t)\rangle 
=\langle\tau +t,\pi|\Psi (0)\rangle 
=\langle 0,\pi|\Psi (t+\tau )\rangle 
\;\;. \end{equation} 
This is the case, for example, if $|\Psi (t)\rangle$ is defined like $|t\rangle$ there, however, 
with some particular initial conditions for the paths contributing to the 
functional integral, or if it is a superposition of such states.    
  
Using this property and shifting and renaming variables, 
the Eq.\,(\ref{discrSchroed}) can be rewritten: 
\begin{equation}\label{discrSchroed1}
\langle 0,\pi'|\Psi (\tau'+T)\rangle 
=\int\mbox{d}\tau\;P(\tau -t) 
\exp [-i(\tau'+T-\tau )\widehat{\cal H}(\pi' ,-i\partial_{\pi'})]
\langle 0,\pi'|\Psi (\tau )\rangle
\;\;. \end{equation}
Despite that a state needs to exist only on discrete values $t$ of 
the physical time, the Eqs.\,(\ref{ttransl}) or (\ref{discrSchroed1}) require the corresponding 
``wave function'' in $\tau ,\pi$-representation 
to be analytically continued to arbitrary real values of the time argument.   

Considering the deterministic limiting case, $P(\tau -t)=\delta (\tau -t)$ we     
obtain directly from Eq.\,(\ref{discrSchroed1}):    
\begin{equation}\label{discrSchroed2}
\langle 0,\pi'|\Psi (\tau'+T)\rangle 
=\exp [-i(\tau'+T-t)\widehat{\cal H}(\pi' ,-i\partial_{\pi'})]
\langle 0,\pi'|\Psi (t)\rangle
\;\;. \end{equation}
Thus, the usual formal solution in terms of an exponentiated Hamilton operator 
of a standard Schr\"odinger equation is recovered, i.e., the Eq.\,(\ref{discrSchroed}) 
here is equivalent to such an equation. 
 
In more general cases, for example, with a Gaussian probability distribution,   
$P(\tau -t)\propto\exp [-\gamma (\tau -t)^2]$, one might suspect that the evolution operator 
on the right-hand side of Eq.\,(\ref{discrSchroed}) or (\ref{discrSchroed1}) is not unitary, 
even if $\widehat{\cal H}$ is hermitean. However, it is easily seen that the Ansatz,  
\begin{equation}\label{solAnsatz} 
\langle\tau ,\pi |\Psi (t)\rangle\equiv
\exp [-i(\tau +t)\widehat{\cal H}(\pi ,-i\partial_{\pi})]\langle 0,\pi |\Psi (0)\rangle\
\;\;, \end{equation} 
solves Eq.\,(\ref{discrSchroed}) also for any normalized 
distribution with $P(\tau ;t)=P(\tau -t)$. In this sense, the Eq.\,(\ref{discrSchroed}) is indeed 
formally equivalent to a Schr\"odinger equation. We will study this in more detail in Section\,6, 
paying attention to the nonstandard form of the effective Hamiltonian (\ref{Hamiltonian}) 
in several examples.  

The above Ansatz suggests to introduce {\it stationary 
states} defined by the following relations:  
\begin{eqnarray}\label{stationary} 
\langle\tau ,\pi|\Psi_n(t)\rangle&\equiv&\exp (-iE_nt)\langle\tau ,\pi|\Psi_n(0)\rangle 
=\exp (-iE_n(t+\tau ))\langle 0,\pi|\Psi_n(0)\rangle
\\ [1ex] \label{stationary1}
&\equiv&\exp (-iE_n(t+\tau ))\langle \pi|\Psi_n\rangle
\;\;, \end{eqnarray} 
in accordance with Eqs.\,(\ref{ttransl}), and assuming: 
\begin{equation}\label{stateq} 
\widehat{\cal H}(\pi ,-i\partial_{\pi})]
\langle 0,\pi |\Psi_n\rangle =E_n\langle 0,\pi |\Psi_n\rangle 
\;\;, \end{equation} 
with a discrete spectrum $\{ E_n\}$, in order to be definite. 

Finally, we remark that there is no $\hbar$ in our 
equations. If introduced, it would merely act as a conversion factor of units. 
On the other hand, there is the intrinsic scale of the clock's unit time interval $T$. The significance of this can be analyzed in a treatment where clock and system are 
part of one universe and interact \cite{E03a}. 
 
Before we will illustrate in some examples the type of Hamiltonians and 
stationary ``wave equations'' that one obtains, 
we have to first address the classical observables and their place in the emergent 
quantum theory. In particular, we need to implement the classical Hamiltonian constraint,  
which is an essential ingredient related to 
the gauge symmetry in a time-reparametrization invariant classical theory.  

\section{Observables}  
It follows from our introduction of states in Section\,3, see particularly Eqs.\,(\ref{Zdef})--(\ref{adjstate}), 
how the classical observables of the underlying mechanical system can be calculated. Considering observables 
which are function(al)s of the phase space variables $\varphi$, the definition of their expectation value 
at physical time $t$ is obvious: 
\begin{eqnarray}\label{expect1} 
\langle O[\varphi ];t\rangle&\equiv&\int\mbox{d}\tau\;P(\tau ;t)O[-i\frac{\delta}{\delta J(\tau )}]\log Z[J]|_{J=0}
\\ [1ex] \label{expect2}
&=&Z^{-1}\int\mbox{d}\tau\mbox{d}\pi\;P(\tau -t)
\langle 0|\tau ,\pi\rangle O[\varphi (\tau )]\langle\tau ,\pi |0\rangle 
\\ [1ex] \label{expect3} 
&=&Z^{-1}\int\mbox{d}\tau\mbox{d}\pi\;P(\tau )
\langle t|\tau ,\pi\rangle O[\varphi (\tau +t)]\langle\tau ,\pi |t\rangle 
\\ [1ex] \label{expect4} 
&=&Z^{-1}\int\mbox{d}\tau\mbox{d}\pi\;P(\tau )
\langle t|\tau ,\pi\rangle O[-i\partial_\pi ]\langle\tau ,\pi |t\rangle 
\\ [1ex] \label{expect5}
&=&\langle\Psi (t)|\widehat O[\varphi ]|\Psi (t)\rangle
\;\;, \end{eqnarray} 
where $Z\equiv Z[0]$, all states refer to $J=0$ as well, and where: 
\begin{equation}\label{phihat}
\widehat O[\varphi ]\equiv O[\widehat\varphi ]\;\;,\;\;\; 
\widehat\varphi\equiv -i\partial_\pi 
\;\;, \end{equation} 
in $\tau ,\pi$-representation. In Eqs.\,(\ref{expect2})--(\ref{expect3}) 
the notation is symbolical, since the observable should be properly included 
in the functional integral defining the ket state, for example. Furthermore, in the last Eq.\,(\ref{expect5}),  
the preceding expression is rewritten for the case of a generic state $|\Psi (t)\rangle$ ($J=0$), 
with the scalar product to be evaluated as in (\ref{expect4}), or as defined in 
Eq.\,(\ref{overlap}) before.    

Thus, a classical observable is represented by the corresponding function(al) of a suitably defined 
{\it momentum} operator. Furthermore, its expectation value at physical time $t$ is represented by the  
effective quantum mechanical expectation value of the corresponding operator with respect to the physical-time 
dependent state under consideration, which incorporates the 
weighted average over the proper times $\tau$, according to the distribution $P$. Not quite surprisingly, 
the evaluation of expectation values involves an integration over the whole $\tau$-parametrized 
``history'' of the states.

Furthermore, making use of the evolution operator $\widehat U$ of Section\,4, in order to refer observables at different 
proper times $\tau_k$ to a common reference point $\tau$, one can construct {\it correlation functions} of 
observables as well, similarly as in Ref.\,\cite{Wetterich02}, for example.   

The most important observable for our present purposes is the classical Hamiltonian, $H(\varphi )$, 
which enters the Hamiltonian constraint of a classical reparametrization invariant system. It is, by 
assumption, a constant of the classical motion. However, it is easy to see that also its quantum 
descendant, $\widehat H(\varphi )\equiv H(\widehat\varphi )$, is conserved, since it commutes 
with the effective Hamiltonian of Eq.\,(\ref{Hamiltonian}): 
\begin{eqnarray}\label{Hconserv} 
[\widehat H,\widehat{\cal H}]&=& 
H(-i\partial_\pi )\;\pi \cdot\omega\cdot
\frac{\partial}{\partial\varphi}H(\varphi )|_{\varphi =-i\partial_{\pi}}
-\pi \cdot\omega\cdot
\frac{\partial}{\partial\varphi}H(\varphi )|_{\varphi =-i\partial_{\pi}}\;H(-i\partial_\pi )
\nonumber \\ [1ex] \label{Hconserv1} 
&=&\frac{\partial}{\partial\varphi}H(\varphi )|_{\varphi =-i\partial_{\pi}} 
\cdot\omega\cdot
\frac{\partial}{\partial\varphi}H(\varphi )|_{\varphi =-i\partial_{\pi}}\;=\;0 
\;\;, \end{eqnarray} 
due to the antisymmetric character of the symplectic matrix. 
Therefore, it suffices to implement the Hamiltonian constraint at an arbitrary  
time. 
  
Then, the constraint of the form $C_H\equiv H(\varphi )-\epsilon\simeq 0$ could be    
incorporated into the definition of the states in Eq.\,(\ref{state}) 
by including an extra factor $\delta (C_H)$ into the functional integral, and 
analogously for the adjoint states. Exponentiating the $\delta$-function, we can  
pull the exponential out of the functional integral, similarly as before.  
Thus, we find the following operator representing the constraint: 
\begin{equation}\label{constraint} 
\widehat{C}\equiv
\int\mbox{d}\lambda\;\exp\Big (i\lambda (\widehat H(\varphi )-\epsilon )\Big )
=\delta (\widehat{C}_H)
\;\;, \end{equation}
which acts on states as a projection operator. It admits in the functional 
integral representing a state only those paths that obey the constraint; in particular, 
see Eq.\,(\ref{state}), it enforces the constraint at the time $\tau +t$.  
Similarly, a projector can be included into the definition of the generating functional,  Eq.\,(\ref{Zdef}), in order to represent the Hamiltonian constraint. 
  
Supplementing Eqs.\,(\ref{expect1})--(\ref{expect5}) by the 
insertion of the Hamiltonian constraint, the properly constrained expection 
values of observables should be calculated according to: 
\begin{equation}\label{Cexpect}  
\langle O[\varphi ];t\rangle_H\equiv\langle\Psi (t)|\widehat O[\varphi ]\widehat{C}|\Psi (t)\rangle
\;\;, \end{equation}
which, in general, will deviate from the results of the previous definition.

Finally, also the eigenvalue problem of stationary 
states, see Eqs.\,(\ref{stationary})--(\ref{stateq}), 
should be studied in the projected subspace: 
\begin{equation}\label{eigenvalue}  
\widehat{\cal H}\widehat{C}|\Psi\rangle =E\widehat{C}|\Psi\rangle 
\;\;, \end{equation} 
to which we shall return in the following section. 

\section{Examples of quantum systems with underlying \\ deterministic dynamics}   
The purpose of the following examples is to illustrate how 
the quantum mechanics works in practice which emerges from various deterministic classical systems 
along the lines presented in Sections\,2--5 before. In particular,  
we solve the stationary Eq.\,(\ref{stateq}) with the effective Hamiltonian $\widehat{\cal H}$ 
of Eq.\,(\ref{Hamiltonian}). 

We shall see, however, that acceptable quantum models with a stable groundstate can only be arrived at in this 
way, if a regularization of the respective Hamiltonian and subsequently a suitable continuum limit are  
performed. In a sense, the spectrum of the Hamiltonian (\ref{Hamiltonian}) is too rich, 
it admits additional unphysical states, which  
have to be eliminated. While this approach covers a large number of models, the meaning of 
and possible restrictions on the ad hoc adopted regularization certainly deserve further study. 
  
Our first example, starting with a ``timeless'' classical harmonic oscillator, 
mainly serves to demonstrate that the present general formalism allows to recover 
results of 't\,Hooft's cellular automaton model \cite{tHooft01}. The second model, employing 
the parameterized classical relativistic particle, is considered 
by many as a testing ground for any techniques developed to deal with reparametrization invariant 
theories in general, like general relativity or string theory. 
It leads to an interacting quantum model, provided we judiciously choose the arbitrary 
phases introduced by the regularization. 

\subsection{Quantum system with classical harmonic oscillator beneath}  
In principle, all {\it integrable models} can be presented as collections of  
harmonic oscillators. Therefore, we begin with  
the harmonic oscillator of unit mass and of frequency $\Omega$. The action is: 
\begin{equation}\label{oscaction} 
S\equiv\int\mbox{d}t\;\Big (\frac{1}{2\lambda}(\partial_tq)^2-\frac{\lambda}{2}(\Omega^2q^2-2\epsilon)\Big ) 
\;\;, \end{equation} 
where $\lambda$ denotes the arbitrary lapse function, i.e. Lagrange multiplier for the Hamiltonian constraint, and $\epsilon >0$ is the parameter fixing the energy presented by this constraint. 
   
Introducing the proper time, $\tau\equiv\int\mbox{d}t\;\lambda$, the Hamiltonian equations of 
motion and Hamiltonian constraint for the oscillator are:  
\begin{eqnarray}\label{osceom1} 
\partial_\tau q=p\;\;,\;\;\;\partial_\tau p=-\Omega^2q\;\;, 
\\ [1ex] \label{oscconstraint} 
\frac{1}{2}(p^2+\Omega^2q^2)-\epsilon =0
\;\;, \end{eqnarray} 
respectively.  
  
Comparing the general structure of the equations of motion, Eq.\,(\ref{eom}), with 
the ones obtained here, we identify the effective Hamilton operator, Eq.\,(\ref{Hamiltonian}), 
while the constraint operator follows from Eq.\,(\ref{constraint}): 
\begin{eqnarray}\label{oscHamiltonian}
&\;&\widehat{\cal H}=-(\pi_q\widehat\varphi_p-\Omega^2\pi_p\widehat\varphi_q)= 
-\pi_q(-i\partial_{\pi_p})+\Omega^2\pi_p(-i\partial_{\pi_q})\;\;, 
\\ [1ex] \label{osccoperator} 
&\;&\widehat{C}
=\delta (\widehat\varphi_p^{\;2}+\Omega^2\widehat\varphi_q^{\;2}-2\epsilon )=
\delta (\partial_{\pi_p}^{\;2}+\Omega^2\partial_{\pi_q}^{\;2}+2\epsilon)
\;\;, \end{eqnarray}
respectively. Here we employ the convenient notation 
$\varphi^a\equiv (\varphi_q;\varphi_p)$, and 
correspondingly $\pi^a\equiv (\pi_q;\pi_p),\;\partial_\pi^a\equiv (\partial_{\pi_q};\partial_{\pi_p})$.    
Further simplifying this with the help of polar coordinates, 
$\pi_q\equiv -\Omega\rho\cos\phi$ and $\pi_p\equiv\rho\sin\phi$, we obtain: 
\begin{eqnarray}\label{angoscHamiltonian}
&\;&\widehat{\cal H}
=\Omega\widehat L_z=-i\Omega\partial_\phi\;\;, 
\\ [1ex] \label{osccoperator1} 
&\;&\widehat{C}
=\delta (\Delta_2+2\epsilon)
=\delta (\partial_\rho^{\;2}+\rho^{-1}\partial_\rho +\rho^{-2}\partial_\phi^{\;2}+2\epsilon )
\;\;, \end{eqnarray}
where $\widehat L_z$ denotes the $z$-component of the usual angular momentum 
operator and $\Delta_2$ the Laplacian in two dimensions. 

We observe that the eigenfunctions of the eigenvalue problem 
posed here factorize into a radial and an angular part. The radial eigenfunction, a  
cylinder function,  is important for the calculation of expectation values of 
certain operators and the overall normalization of the resulting wave functions. 
However, it does not influence the most interesting spectrum of the Hamiltonian.     
  
In order to proceed, we discretize the angular derivative, Eq.\,(\ref{angoscHamiltonian}). 
In the absence of the full angular momentum algebra, we would otherwise encounter a discrete 
yet unbound spectrum, lacking a groundstate.  
 
While we will mostly choose to work with an asymmetric discretization 
(\underline{Case A}), we will here also show the symmetric discretization 
(\underline{Case B}), in order to appreciate 
the differences, if any. In any case, the spectrum should and will turn out to be independent 
of this choice in the continuum limit. 

\underline{Case A}. Here, the energy eigenvalue problem consists in: 
\begin{equation}\label{osceigen}
\widehat{\cal H}\psi (\phi_n) 
=-i(\Omega N/2\pi )\Big (\psi (\phi_{n+1})-\psi (\phi_n)\Big ) =E\psi (\phi_n) 
\;\;, \end{equation} 
with $\phi_n\equiv 2\pi n/N$, $1\leq n\leq N$, and the continuum limit will be 
considered momentarily.   
  
The complete orthonormal set of eigenfunctions and the eigenvalues are easily found: 
\begin{eqnarray}\label{osceigenfunction} 
\psi_m(\phi_n)&=&N^{-1/2}\exp [i(m+\delta )\phi_n]\;\;,\;\;1\leq m\leq N  
\;\;, \\ [1ex] \label{osceigenvalue} 
E_m&=&i(\Omega N/2\pi )\Big (1-\exp [2\pi i(m+\delta)/N]\Big )
\\ [1ex] \label{osceigenvalue1}
&\stackrel{N\rightarrow\infty}{\longrightarrow}&\Omega (m+\delta )\;\;,\;\;m\in\mathbf{N}  
\;\;, \end{eqnarray} 
where $\delta$ is an arbitrary real constant. Next, we turn to 
the symmetric discretization.   

\underline{Case B}.
Here we have instead: 
\begin{equation}\label{osceigensym}
\widehat{\cal H}\psi (\phi_n) 
=-i(\Omega N/4\pi )\Big (\psi (\phi_{n+1})-\psi (\phi_{n-1})\Big ) =E\psi (\phi_n) 
\;\;, \end{equation} 
with $\phi_n$ as before. This is solved by the same eigenfunctions as before, Eq.\,(\ref{osceigenfunction}). 
However, in this case the eigenvalues are real even before taking the continuum limit: 
\begin{equation}\label{osceigenvaluesym}
E_m=(\Omega N/2\pi )\sin [2\pi (m+\delta )/N]
\;\stackrel{N\rightarrow\infty}{\longrightarrow}\;\Omega (m+\delta )\;\;,\;\;m\in\mathbf{N}  
\;\;. \end{equation} 
Thus, we learn that the spectrum in the continuum limit is indeed real and independent of the regularizations employed to suitably define the Hamiltonian, as it should be.

The freedom to choose the constant $\delta$, which arises from 
the regularization of the Hamilton operator, is most wellcome. Choosing $\delta\equiv -1/2$,  
we arrive at the {\it quantum harmonic oscillator}, having started from a corresponding 
classical system here. Thus, we recover in a straightforward way 't Hooft's result, derived from a cellular automaton \cite{tHooft01}. See also Ref.\,\cite{E03} for the  
completion of a similar quantum model. In the following example we will encounter 
one more model of this kind and demonstrate its solution in detail. 

Also in the following example the regularized eigenvalues are {\it complex} and the real 
spectrum only emerges in the continuum limit. Again, this is due to the fact that we choose to discretize first-order derivatives asymmetrically, and could be avoided as shown. 

\subsection{Quantum system with classical relativistic particle beneath}  
Introducing proper time similarly as in Ref.\,\cite{E03}, the equations of motion and 
the Hamiltonian constraint of the reparametrization invariant kinematics of a 
classical relativistic particle of mass $m$ are given by: 
\begin{eqnarray}\label{releom}
\partial_\tau q^\mu =m^{-1}p^\mu\;\;,\;\;\;\partial_\tau p^\mu =0
\;\;, \\ [1ex] \label{relc} 
p\cdot p-m^2=0 
\;\;, \end{eqnarray} 
respectively. Here we have $\varphi^a\equiv (q^0,\dots ,q^3;p^0,\dots ,p^3),\;a=1,\dots ,8$; 
four-vector products involve the Minkowski metric, 
$g^{\mu\nu}\equiv\mbox{diag}(1,-1,-1,-1)$. 
   
Proceeding as before, we identify the effective Hamilton operator: 
\begin{equation}\label{relHamiltonian}
\widehat{\cal H}=-m^{-1}\pi_q\cdot\widehat\varphi_p=-m^{-1}\pi_q\cdot(-i\partial_{\pi_p}) 
\;\;, \end{equation} 
corresponding to Eq.\,(\ref{Hamiltonian}); the notation is as   
introduced after Eq.\,(\ref{osccoperator}), however, involving four-vectors.  
Furthermore, the Hamiltonian constraint 
is represented by the operator: 
\begin{equation}\label{relconstraint} 
\widehat{C}
=\delta (\widehat\varphi_p^{\;2}-m^2)=\delta (\partial_{\pi_p}^{\;2}+m^2)
\;\;, \end{equation}
according to Eqs.\,(\ref{constraint}) and (\ref{relc}). 
After a Fourier transformation, which replaces the variable $\pi_q$ by a derivative (four-vector) $+i\partial_x$, 
and renaming $\pi_p\equiv\bar x$, we obtain: 
\begin{equation}\label{Hplusc}  
\widehat{\cal H}=-m^{-1}\partial_x\cdot\partial_{\bar x}\;\;,\;\;\;
\widehat{C}=\delta (\partial_{\bar x}^{\;2}+m^2) 
\;\;, \end{equation} 
i.e., a more transparent form of the Hamilton and constraint operators, respectively. 

Before embarking to further analyze this model, some general remarks seem in order here.  
It is well known from ordinary quantum (field) theory that the harmonic oscillator is peculiar in 
many respects. Therefore, the reader should not be misled by the results of Section\,6.1, 
where we obtain a quantum harmonic oscillator spectrum from an underlying classical harmonic oscillator model. In particular, it may appear as if we have invented just one more quantization method, in line with quantization via canonical commutators, stochastic quantization, etc. However, we stress that this is {\it not} the case \cite{tHooft01,Smolin}. 

It seems an accident of the harmonic system 
that the usual quantized energy spectrum results here. 
This is revealed by the fact, already demonstrated in Refs.\,\cite{tHooft01,ES02,E03}, 
that localization with respect to the coordinate $q$ of the underlying classical model 
has nothing to do with localization with respect to the operator $\hat q$, which is  
introduced a posteriori when interpreting the emergent quantum Hamiltonian corresponding 
to said spectrum. Rather, such localized quantum (oscillator) states are widely spread over 
the q-space of the underlying classical model. 
We will encounter such operators in the following example again. Furthermore, 
the usual $\hat p,\hat q$-commutator algebra obtains corrections here, as long as the 
regularization is not removed. 
  
Therefore, generally, we do not expect to find the usual 
quantized counterpart of a classical reparametrization 
invariant model in the present approach based on discrete physical time. 
There will not be the usual one-to-one 
correspondence. To put it differently, the {\it classical limit} of emergent quantum 
theories {\it cannot} be expected to give back the {\it underlying} classical model.   
Further general remarks in this vein may be found in Ref.\,\cite{tHooft01}.  

The following 
discussion of the free relativistic particle should be seen in this light. While any standard 
quantization method produces an unbound spectrum with notorious negative 
energy states, we show in detail here that careful application of the regularization 
indeed allows to produce an acceptable quantum model 
in the continuum limit, which is different from the underlying classical 
model. 

The eigenvalue problem is 
solved after discretizing the system with a hypercubic 
(phase space) lattice of volume $L^8$ 
(lattice spacing $l\equiv L/N$) 
and periodic boundary conditions, 
for example. Similarly as in the oscillator case, we here obtain the eigenfunctions:
\begin{equation}\label{releigenfunction}   
\psi_{k_x,k_{\bar x}}(x_n,\bar x_n)=N^{-1}
\exp [i(k_x+\delta_x)\cdot x_n+i(k_{\bar x}+\delta_{\bar x})\cdot \bar x_n] 
\;\;, \end{equation} 
with coordinates $x_n^\mu\equiv ln^\mu$ and   
momenta $k_x^\mu\equiv 2\pi k^\mu /L$, with $1\leq n^\mu ,k^\mu\leq N$, 
and where $\delta_x^\mu$ are arbitrary real constants,  
for all $\mu =0,\dots ,3$ (analogously $\bar x_n^\mu$, $k_{\bar x}^\mu$, 
$\delta_{\bar x}^\mu$). 

These are the eigenfunctions, which are of plane-wave type, of the stationary  
Schr\"odinger equation, $\widehat{\cal H}\widehat C\psi =E\widehat C\psi$, 
with $\widehat{\cal H}$ 
and $\widehat C$ 
from Eqs.\,(\ref{Hplusc}) discretized analogously to \underline{Case A} of Section\,6.1. 
The four-vector momenta $k_x^\mu,k_{\bar x}^\mu$ label different eigenfunctions, 
a familiar feature in quantum mechanics, and the constraint will be implemented shortly. 
Note that the phases 
$\delta_x^\mu,\delta_{\bar x}^\mu$ can still be chosen arbitrarily; 
however, since they must be fixed once for all, they cannot possibly absorb the 
variable four-momenta. Furthermore, we emphasize that the phase space lattice has been 
introduced for the only purpose of regularizing the Hamiltonian, to be followed by a 
suitable continuum limit. At present, however, it is unrelated to the underlying 
discreteness of physical time.        

The corresponding energy eigenvalues are: 
\begin{eqnarray}\label{releigenvalue}  
E_{k_x,k_{\bar x}}&=&
-m^{-1}l^{-2}\Big ((\exp [il(k_x+\delta_x)^0]-1)(\exp [il(k_{\bar x}+\delta_{\bar x})^0]-1)
\\ 
&\;&\;\;\;\;\;\;\;\;\;\;-\sum_{j=1}^3(\exp [il(k_x+\delta_x)^j]-1)(\exp [il(k_{\bar x}+\delta_{\bar x})^j]-1)\Big )
\nonumber \\ [1ex] \label{releigenvalue1}
&=&
m^{-1}(k_x+\delta_x)\cdot (k_{\bar x}+\delta_{\bar x})+O(l)
\;\;, \end{eqnarray} 
where is $L$ is kept constant in the continuum limit, $l\rightarrow 0$; 
again, the four-momenta $k_x^\mu,k_{\bar x}^\mu$ label and 
determine the energies. Furthermore, in this limit, 
one finds that the Hamiltonian constraint requires timelike ``on-shell'' vectors 
$k_{\bar x}$, obeying $(k_{\bar x}+\delta_{\bar x})^2=m^2$, 
while leaving $k_x$ unconstrained.   
  
Continuing, we perform also the infinite volume limit, $L\rightarrow\infty$, 
which results in a continuous energy spectrum in Eq.\,(\ref{releigenvalue1}). 
We observe that no matter how we choose the constants $\delta_x,\delta_{\bar x}$, 
the spectrum will not be positive definite. Thus, the emergent model appears not 
to be acceptable, since it does not lead to a stable groundstate. 
  
However, let us proceed more carefully with the various limits involved 
and show that indeed a well-defined quantum model is obtained.  
For simplicity, considering (1+1)-dimensional Minkowski space and anticipating 
the massless limit, we rewrite Eq.\,(\ref{releigenvalue1}) explicitly: 
\begin{equation}\label{releigen}
E_{k,\bar k}=-(\frac{2\pi}{\sqrt mL})^2
(\bar k^1+\bar\delta^1)\Big ((k^0+\delta^0)+(k^1+\delta^1)\Big )+O(m)
\;\;, \end{equation} 
where we suitably rescaled and renamed the constants and the momenta, which run in 
the range $1\leq\bar k^1,k^{0,1}\leq N\equiv 2s+1$. Furthermore, we incorporated the 
Hamiltonian (on-shell) constraint, such 
that only the positive root contributes: 
$\bar k^0+\bar\delta^0=|\bar k^1+\bar\delta^1|+O(m^2)=-(\bar k^1+\bar\delta^1)+O(m^2)$. 
This can be achieved by suitably choosing $\bar\delta^{0,1}$.     

In fact, just as in the previous harmonic oscillator case, 
the choice of the phase constants is crucial in determining the quantum model. 
Here we set: 
\begin{equation}\label{deltas}
\bar\delta^0\equiv\frac{1}{2}\;,\;\;\bar\delta^1\equiv \frac{1}{2}-2s-3\;,\;\;
\delta^{0,1}\equiv 0
\;\;. \end{equation} 
We observe that this choice implies that in the continuum limit, 
with $l\rightarrow 0$, since $N\rightarrow\infty$, we have $s\rightarrow\infty$ and thus 
$\bar\delta^1\rightarrow -\infty$.  
Incorporating these phases, we obtain the manifestly positive definite spectrum: 
\begin{equation}\label{releigen1}
E(\bar s_z,s_z^{0,1})=(\frac{2\pi}{\sqrt mL})^2\Big ((\bar s_z+s+\frac{1}{2})+1\Big )  
\Big ((s_z^0+s+\frac{1}{2})+(s_z^1+s+\frac{1}{2})+1\Big )+O(m)
\;\;, \end{equation}
where also the (half)integer quantum numbers $\bar s_z,s_z^{0,1}$ are introduced,  
all in the range $-s\leq s_z\leq s$, which correctly replace $\bar k^1,k^{0,1}$. 

Recalling the algebra of the $SU(2)$ generators, with $S_z|s_z\rangle =s_z|s_z\rangle$ 
in particular, we are led to consider the generic operator: 
\begin{equation}\label{Hdiag} 
h\equiv S_z+s+\frac{1}{2}   
\;\;, \end{equation}
i.e., diagonal with respect to $|s_z\rangle$-states of the (half)integer representations determined by $s$.  
In terms of such operators, we obtain the regularized Hamiltonian corresponding to Eq.\,(\ref{releigen1}): 
\begin{equation}\label{relHreg}
\widehat{\cal H}=
(\frac{2\pi}{\sqrt mL})^2\Big (1+\bar h+h_0+h_1+\bar h(h_0+h_1)\Big )+O(m)
\;\;, \end{equation}
which will turn out to represent three coupled harmonic oscillators,   
including an additional contribution to the vacuum energy. 

A Hamiltonian of the type of h has been the starting point of 't Hooft's analysis \cite{tHooft01}, 
which we adapt for our purposes in the following.  

Continuing with standard notation, we have $S^2\equiv S_x^{\;2}+S_y^{\;2}+S_z^{\;2}=s(s+1)$, which suffices 
to obtain the following identity: 
\begin{equation}\label{Hsq} 
h=\frac{1}{2s+1}\Big (S_x^{\;2}+S_y^{\;2}+\frac{1}{4}+h^2\Big ) 
\;\;. \end{equation} 
Furthermore, using $S_\pm\equiv S_x\pm iS_y$, we define coordinate and conjugate 
momentum operators: 
\begin{equation}\label{qp}
\hat q\equiv\frac{1}{2}(aS_-+a^\ast S_+)\;\;,\;\;\; 
\hat p\equiv\frac{1}{2}(bS_-+b^\ast S_+)
\;\;, \end{equation} 
where $a$ and $b$ are complex coefficients. Calculating the basic commutator with the help 
of $[S_+,S_-]=2S_z$ and using Eq.\,(\ref{Hdiag}), we obtain: 
\begin{equation}\label{commutator} 
[\hat q,\hat p]=i(1-\frac{2}{2s+1}h)
\;\;, \end{equation} 
provided we set $\Im (a^\ast b)\equiv -2/(2s+1)$. Incorporating this,  
we calculate: 
\begin{equation}\label{sumsqu}
S_x^{\;2}+S_y^{\;2}=\frac{(2s+1)^2}{4}\left (
|a|^2\hat p^2+|b|^2\hat q^2-(\Im a\cdot\Im b+\Re a\cdot\Re b)\{ \hat q,\hat p\}\right )
\;\;. \end{equation}  
In order to obtain a reasonable Hamiltonian in the continuum limit,  
we set: 
\begin{equation}\label{ab} 
a\equiv i\frac{\Omega^{-1/2}}{\sqrt{s+1/2}}\;\;,\;\;\;b\equiv\frac{\Omega^{1/2}}{\sqrt{s+1/2}} 
\;\;,\;\;\;\Omega\equiv (\frac{2\pi}{\sqrt mL})^2
\;\;. \end{equation}  
Then, the previous Eq.\,(\ref{Hsq}) becomes: 
\begin{equation}\label{Hsq1} 
\Omega h=\frac{1}{2}\hat p^2+\frac{1}{2}\Omega^2\hat q^2
+\frac{1}{(2s+1)\Omega}\Big (\frac{1}{4}\Omega^2+(\Omega h)^2\Big )
\;\;, \end{equation}
reveiling a nonlinearly modified harmonic oscillator Hamiltonian, similarly as in Ref.\,\cite{E03}. 

Now it is safe to consider the continuum limit, $2s+1=N\rightarrow\infty$, 
keeping $\sqrt mL$ and $\Omega$ finite. This produces the usual 
$\hat q,\hat p$-commutator in Eq.\,(\ref{commutator}) for states with limited energy 
and the standard harmonic oscillator Hamiltonian in Eq.\,(\ref{Hsq1}). 

Using these results in Eq.\,(\ref{relHreg}), the Hamilton operator of the  
emergent quantum model is obtained:  
\begin{equation}\label{relHreg1}
\widehat{\cal H}=\Omega +\frac{1}{2}\sum_{j=\bar 1,0,1}\Big (\hat p_j^{\;2}+\Omega^2\hat q_j^{\;2}\Big )
+\frac{1}{4\Omega}(\hat p_{\bar 1}^{\;2}+\Omega^2\hat q_{\bar 1}^{\;2})
\sum_{j=0,1}\Big (\hat p_j^{\;2}+\Omega^2\hat q_j^{\;2}\Big )
\;\;, \end{equation}
where the massless limit together with the infinite volume limit is carried out, 
$m\rightarrow 0,\;L\rightarrow\infty$, in such a way that $\Omega$ remains finite.  

The resulting Hamiltonian here is well defined in terms of continuous operators 
$\hat q$ and $\hat p$, as usual, and has a positive definite spectrum. The coupling term might 
appear somewhat more familiar, if the oscillator algebra is realized in terms of 
bosonic creation and annihilation operators. 

In Refs.\,\cite{ES02,E03} we calculated 
the matrix elements of operators $\hat q,\hat p$ with respect to the SU(2) basis 
of primordial states in an analogous case, showing that {\it localization of the quantum 
oscillator has little to do with localization in the classical model beneath}, as 
we mentioned before. 

Finally, we remark that had we chosen $\bar\delta^{0,1}=\delta^{0,1}\equiv 1/2$, 
instead of Eqs.\,(\ref{deltas}), 
then a relative sign between terms would remain, originating from the 
Minkowski metric, and this would yield the Hamiltonian 
$\widehat{\cal H}\propto (1+\bar h)(h_0-h_1)$, which is not 
positive definite. Similarly, any 
symmetric choice, $\bar\delta^{0,1}=\delta^{0,1}\equiv\delta$ would 
suffer from this problem.     
 
This raises the important 
issue of the role of canonical transformations, and of symmetries in particular.  
It is conceivable that symmetries will play a role in restricting the apparent 
arbitrariness of the regularization. Independence from the choice of phase space coordinates 
employed in actual calculations might be a desirable feature. A preliminary study of an anharmonic 
oscillator indicates that the simple discretization used here possibly needs to be  
improved, in order to fullfill this.    

\subsection{Remarks}  
Concluding this section, we may state that the features which we illustrated in the  
previous examples, especially the necessity of regularization (discretization), 
promise to make genuinely interacting models quite difficult to analyze.  
Interesting results may perhaps be found with the help of spectrum generating algebras and/or 
some to-be-developed perturbative methods.      

We find it interesting that our general Hamilton operator, Eq.\,(\ref{Hamiltonian}), 
does not allow for the direct addition of a constant energy term, while in its regularized  
form this is possible, due to the appearance of an arbitrary phase. Choosing the latter 
determines the groundstate energy, which survives the continuum limit.     

Leaving aside the Hamiltonian constraint momentarily, we observe that the Hamilton operator equation, also in cases 
with large numbers $n$ of coupled degrees of freedom, amounts to systems of    
first order quasi-linear partial differential equations. They  
can be studied by the method of characteristics \cite{CH}.  
Thus, one finds one inhomogeneous equation, which can be trivially integrated. 
Furthermore, the remaining $2n$ 
equations for the characteristics present nothing but the classical Hamiltonian 
equations of motion.   
  
Therefore, integrable classical models can be decoupled  
at the level of the characteristic equations by canonical transformations,   
if they can be applied freely at the pre-quantum level. 
Classical crystal-like models with harmonic forces, or free field theories, respectively, 
will thus give rise to corresponding free quantum mechanical systems. These are  
constructed in a different way in Refs.\,\cite{tHooft01}. Presumably, the 
(fixing of a large class of)  
gauge transformations invoked there can be related to the existence of integrals of 
motion implied by integrability here. 

Finally, we point out that the Hamiltonian equations of motion preclude motion into  
classically forbidden regions of the underlying system. Nevertheless, {\it quantum mechanical tunneling} is an intrinsic property of the quantum oscillator models that we obtained. 
 
\section{Conclusions} 
In the present paper, we pursue the view that quantum mechanics is an emergent description of nature, 
which possibly can be based on classical, pre-quantum concepts.  

Our approach is motivated by the ongoing construction of a reparametrization-invariant time. 
In turn, this is based on the observation that ``time passes'' when there is an observable change, which is 
localized with the observer. More precisely, necessary are incidents, i.e. observable unit changes, which are 
recorded, and from which invariant quantities characterizing the change of the evolving system can be derived \cite{ES02,E03}. 
  
Presently, this has led us to assume the relation between the constructed 
physical time $t$ and standard proper time $\tau$ of the evolving system in the form of 
a statistical distribution,     
$P(\tau ;t)=P(\tau -t)$, cf. Eq.\,(\ref{P}). We assume that the distribution is not 
explicitly time-dependent, which means, the physical clock is practically decoupled 
from the system under study. 
We explore the consequences of this situation for the description of the system. 
  
We have shown how to introduce ``states'', eventually building up a Hilbert space, in terms of certain 
functional integrals, Eqs.\,(\ref{state})--(\ref{adjstate}), which arise from the study of a  
classical generating functional. The latter was introduced earlier in a 
different context, studying classical mechanics in functional form \cite{GRT,GR00}. We employ this  
as a convenient tool, and modify it, in order to describe the observables of reparametrization-invariant systems 
with discrete time (Section\,5). 
Studying the evolution of the states in general (Section\,4), we are led to the Schr\"odinger equation,  
Eq.\,(\ref{discrSchroed}). However, the Hamilton operator, Eq.\,(\ref{Hamiltonian}), has a non-standard form.   
  
We demonstrate that proper regularization of the continuum Hamilton  
operator is indispensable, in order to find a stable groundstate. Limitations 
imposed by symmetries or further constraints and consistency of the procedure need to be clarified.  
Other possible regularization schemes need to be explored. 

Coming back to the probabilistic relation between physical time and the evolution parameter 
figuring in the parameterized classical equations of motion: 
One would like to include the clock degrees of freedom consistently 
into the dynamics, in order to address the closed universe. This can be achieved 
by introducing suitable projectors into the generating functional \cite{Jonathan,E03a}. 
Their task is to represent    
quasi-local detectors which respond to a particle trajectory passing 
through in Yes/No fashion. In a more general setting, such a detector/projector has to 
be defined in terms of observables of the closed system. In this way, typical conditional probabilities 
can be handled, such as describing ``What is the probability of observable $X$ having a value 
in a range $x$ to $x+\delta x$, {\it when} observable $Y$ has value $y$?''. Criteria for selecting the 
to-be-clock degrees of freedom are still unknown, other than simplicity. Most   
likely the resulting description of evolution and implicit notion of physical time will correspond to 
our distribution $P(\tau ;t)$ of Eq.\,(\ref{P}), however, evolving with the system. 
Dissipative and memory effects, which will arise after integration over clock degrees of 
freedom, might play a crucial role in deriving a 
unique large-scale quantum model.       

This stroboscopic quantization emerging from underlying classical dynamics 
certainly may be questioned in many respects. It might violate one or the other assumption of existing 
no-go theorems relating to hidden variables theories. However, we believe it is 
interesting to get closer to a working example, before discussing this. Unitary evolution and   
tunneling effects are recovered in this framework, under the 
proviso of regularization of the continuum formalism. 
 
\subsection*{Acknowledgements} 
It is a pleasure to thank Christof Wetterich for discussion  
and the members of the Institut f\"ur Theoretische Physik (Heidelberg) 
for their kind hospitality, while this work was begun. Correspondence from 
Ennio Gozzi is thankfully acknowledged. This work has been   
supported by CNPq (Brasilia) 690138/02-4 and DAAD (Bonn) A/03/17806. 



\end{document}